\documentclass[aps,showpacs,pre,twocolumn,superscriptaddress,floatfix]{revtex4-1}
\usepackage{graphicx,bm,amssymb,amsmath,dcolumn,hyperref}
\usepackage{multirow}
\usepackage{color}
\usepackage{subfigure}  
\usepackage{epstopdf}
\usepackage{bbm}
\usepackage{graphicx}
\usepackage{hyperref}
\usepackage{threeparttable}
\usepackage{amsthm,enumitem,mathrsfs}
\usepackage[marginal]{footmisc}
\usepackage{array}
\usepackage{tabularx}
\usepackage{makecell}

\newcommand{\scrZ}{{\mathcal Z}}
\newcommand{\scrU}{{\mathcal U}}

\newcommand{\scrF}{{\mathcal F}}
\newcommand{\scrN}{{\mathcal N}}
\newcommand{\scrS}{{\mathcal S}}

\newcommand{\PP}{{\mathbb P}}

\newcommand{\dc}{{d_c}}
\newcommand{\dpp}{{d_p}}
\newcommand{\Kc}{{K_c}}
\newcommand{\dLone}{{d_\textsc{l1}}}
\newcommand{\dLtwo}{{d_\textsc{l2}}}
\newcommand{\dFone}{{d_\textsc{f1}}}
\newcommand{\dFtwo}{{d_\textsc{f2}}}

\begin{document}
	
	\title{Finite-Size Scaling of the High-Dimensional Ising Model in the Loop Representation}
	\date{\today}
	
	\author{Tianning Xiao}
	\thanks{These two authors contributed equally to this paper.}
	\affiliation{Hefei National Research Center for Physical Sciences at the Microscale, University of Science and Technology of China, Hefei 230026, China}
	\author{Zhiyi Li}
	\thanks{These two authors contributed equally to this paper.}
	\affiliation{Department of Modern Physics, University of Science and Technology of China, Hefei, Anhui 230026, China}
	\author{Zongzheng Zhou}
    \email{eric.zhou@monash.edu}
	\affiliation{School of Mathematics, Monash University, Clayton, Victoria 3800, Australia}
	\author{Sheng Fang}
	\email{fs4008@mail.ustc.edu.cn}
	\affiliation{Hefei National Research Center for Physical Sciences at the Microscale, University of Science and Technology of China, Hefei 230026, China}
	\author{Youjin Deng}
	\email{yjdeng@ustc.edu.cn}
	\affiliation{Hefei National Research Center for Physical Sciences at the Microscale, University of Science and Technology of China, Hefei 230026, China}
	\affiliation{Department of Modern Physics, University of Science and Technology of China, Hefei, Anhui 230026, China}
 \affiliation{Hefei National Laboratory,
University of Science and Technology of China, Hefei, Anhui 230088, China}

	\begin{abstract}
		Besides its original spin representation, the Ising model is known to have the Fortuin-Kasteleyn (FK) bond and loop representations, of which the former was recently shown to exhibit two upper critical dimensions $(d_c=4,d_p=6)$. 
		Using a lifted worm algorithm, we determine the critical coupling as $K_c = 0.077\,708\,91(4)$ for $d=7$, which significantly improves over the previous results, and then study critical geometric properties of the loop-Ising clusters on tori for spatial dimensions $d=5$ to 7.
		We show that, as the spin representation, the loop Ising model has only one upper critical dimension at $d_c=4$. 
		However, sophisticated finite-size scaling (FSS) behaviors, like two length scales, two configuration sectors and two scaling windows, still exist as the interplay effect of the Gaussian fixed point and complete-graph asymptotics. Moreover, using the Loop-Cluster algorithm, we provide an intuitive understanding of the emergence of the percolation-like upper critical dimension $d_p=6$ in the FK-Ising model. As a consequence, a unified physical picture is established for the FSS behaviors in all the three representations of the Ising model above $d_c=4$.
	\end{abstract}
	
	
	\maketitle
	
	
	\section{Introduction}
	\label{sec:intro}
		The Ising model is one of the most important models in statistical physics and has wide applications in almost every branch of modern physics~\cite{FriedliVelenik2017,intro_Ising}. 
	Given a lattice $G= (V, E)$ with the vertex set $V$ and edge set $E$, the partition function of the ferromagnetic Ising model with zero field can be written as 
	\begin{equation}
		\mathcal Z_{\rm {spin}}= \sum_{s\in \{-1,1\}^V}  e^{K \sum_{ij\in E }s_i s_j}, 	
		\label{eq:Ising_PF}
	\end{equation}
	where $s_i = \pm 1$ represents spin orientation up or down on vertex $i$, and the coupling strength $K>0$ is proportional to the inverse temperature.
	For the Ising model, there is an upper critical dimension $\dc=4$, above which the critical behavior is governed by the mean-field theory. 
	Two typical approaches of the mean-field solution are the Gaussian fixed point (GFP) solution and the complete graph (CG) solution. 
	The GFP solution is well established in the framework of renormalization group (RG) theory, and gives the RG exponents as  $(y_t, y_h) = (2, 1+d/2)$~\cite{Fernandez2013Random}. 
	The CG is a fully connected and finite graph with all vertexes adjacent to each other. It focuses on finite-size scaling (FSS) behavior and gives effective  RG exponents $(y_t^*, y_h^*) = (d/2, 3d/4)$~\cite{Luijten1997Interaction}.

Besides the spin representation, the Ising model can be reformulated in geometric representations via high-temperature expansion techniques, including the Fortuin-Kasteleyn (FK) bond representation~\cite{grimmett2006random} and loop representation (also known as the random-current representation)~\cite{Parisi1988Statistical}. The FK representation of the Ising model is the $q=2$ case of the general $q$-state random cluster model which is defined as follows. Given a graph $(V, E)$, each edge is either empty or occupied by a bond. Each occupied bond has a statistical weight (relative to the empty one) as $v$, and the fugacity of each connected component (also called cluster) is $q$. The partition function of the random cluster model then reads as
	\begin{equation}
		\mathcal Z_{\rm {FK}}= \sum_{A \subseteq E} q^{k(A)}v^{|A|}, 	
		\label{eq:Ising_FKPF}
	\end{equation}
	where $k(A)$ is the number of clusters and $|A|$ is the number of bonds. For $q=2$, the bond weight $v=e^{2K}-1$.
	The partition function of the loop representation of the Ising model is
	\begin{equation}
		\scrZ_{\rm Loop} = \sum_{A \subseteq E} (\tanh K)^{|A|} {\mathbbm{1}}(A \ \text{is even}) \;.   
	\end{equation}
	Thus, in the loop representation, the weight of an occupied bond is $\tanh K$. The indicator function above means that all vertices on $(V, A)$ must have even degree. These two graphical models can be mapped onto each other through the Loop-Cluster(LC) joint model~\cite{Zhang2020Loop}. This means FK configurations can be generated by placing bonds with probability $\tanh K$ onto the empty edges of loop configurations; and this process is called the LC transformation.

  	\begin{table*}[t]
		\renewcommand\arraystretch{1.5} 
		\centering
		\begin{tabular}{|c|c|c|c|c|}
   \hline 
           & $d$ &  Two length scales  & The vanishing special sector & Two scaling windows\\
			\hline 
 \thead[c]{ \\ \\ \\ FK \\ representation }  &$4<d<6$   & \thead[l]{~Giant cluster: $C_1 \sim L^{y^*_h} \sim R_1^{y^*_h}$ \\   Other clusters: $s \sim R^{y_h}$, $n(s,L) \sim s^{-(1+d/y_h)}$ }    & \thead[l]{Vanishing rate: $P \sim L^{y_h-y_h^*}$ \\ ~~~In the sector: $s \sim R^{y_h}$ }  & \thead[l]{ Width: $O(L^{-y^*_t})$ \\ \qquad  ~~~~~$O(L^{-y_t})$ }\\
     \cline{2-5}
 \thead[c]{ }    & $ d \ge 6$& \thead[l]{~Giant cluster: $C_1 \sim L^{y^*_h} \sim R_1^{9/2}$ \\ Other clusters: $s \sim R^4$, $n(s,L) \sim s^{-(1+d/y_{h,\textsc{p}}^{*})}$} &  \thead[l]{Vanishing rate: $P \sim L^{y_{h,\textsc{p}}^{*}-y_h^*}$ \\ ~~~In the sector: $s \sim R^{4}$ }   &  \thead[l]{Width:  $O(L^{-y^*_t})$ \\  \qquad ~ $\begin{cases}
	                O(L^{-y_{t,\textsc{p}}^{*}}) \\ 
                    O(L^{-y_t})
 	            \end{cases} $ 
              }  \\
  \hline 
 \thead[c]{Loop  \\ representation}  &  $d>4$ & \thead[l]{Giant clusters: $n(s,L)\sim L^{-d} s^{-1}$ \\ \qquad \qquad \qquad ~~ $F_{1,2} \sim L^{y_t^*} \sim R_{1,2}^{y_t}$    \\ Other clusters: $n(s,L) \sim s^{-(1+d/y_t)}$} & \thead[l]{Vanishing rate: $P \sim L^{y_h-y_h^*}$ \\ ~~~In the sector: $F_{1,2} \sim L^{y_t} \sim R_{1,2}^{y_t}$} & \thead[l]{Width: $O(L^{-y^*_t})$ \\ \qquad ~~~~ $O(L^{-y_t})$}     \\ 
  \hline 
  \thead[c]{ Values of \\ RG exponents}  & \multicolumn{4}{c|}{ $(y_t, y_h) = (2, 1+d/2)$, \qquad  $(y_t^*, y_h^*) = (d/2, 3d/4)$, \qquad $(y_{t,\textsc{p}}^{*}, y_{h,\textsc{p}}^{*}) = (d/3, 2d/3)$ }\\
 \hline 
		\end{tabular}
		\caption{Summaries of the two-length-scale behavior, two configuration sectors, and two scaling windows of the Ising model in the FK representation and loop representation on high-d tori. The GFP exponents for the Ising model and percolation are $(y_t, y_h) = (2, 1+d/2)$. The RG exponents for the CG Ising model are $(y^*_t, y^*_h) = (d/2, 3d/4)$ and for the CG percolation model are $(y_{t,\textsc{p}}^{*}, y_{h,\textsc{p}}^{*}) = (d/3, 2d/3)$. On high-d tori ($d\ge6$), the thermodynamic fractal dimension of critical percolation clusters is $d_\textsc{f}=4$. We denote $s$ and $R$ for the size and radius of gyration for a generic cluster, $C_1, F_i$ for the size of the largest FK cluster and for the $i$th-largest loop cluster, $R_1, R_2$ for the radius of gyration of the largest and second largest FK or loop clusters. The Giant cluster(s) represents the largest FK cluster and loop clusters with size of order $L^{y_t^*}$ for the FK and loop representations, respectively. We denote $n(s, L)$ the cluster-number density. For the both FK and loop representations, there is a special sector in the configuration space, the proportion (denote as $P$) of which vanishes asymptotically. For $d\geq 6$, the FK Ising model has a CG-Ising scaling window and a high-d percolation scaling window; the latter consists of a CG-percolation window and a GFP-percolation window. These novel properties suggest that, the FK Ising model has two upper critical dimensions (4 and 6), while the loop Ising model has only one upper critical dimension 4.}  
		\label{tab:FK_two_upper_critical_dimension}
	\end{table*}
	Geometric representations of the Ising model offer several advantages.
	Firstly, they provide a platform for designing efficient Monte Carlo (MC) algorithms, such as the worm algorithm based on the loop representation~\cite{worm_2001}, the Swendsen-Wang algorithm utilizing the FK representation~\cite{Swendsen1987Nonuniversal}, and the LC algorithm derived from the LC joint model~\cite{Zhang2020Loop}. These algorithms enhance computational efficiency and facilitate simulations of the Ising model. 
	Secondly, geometric representations play a significant role in conformal field theory~\cite{Francesco2012Conformal} and stochastic Loewner evolution~\cite{Cardy2005Sle,Kager2004Guide}, enabling a deeper understanding of the spin Ising model. Notably, using the random-current representation, it was proved that the three-dimensional (3D) Ising model exhibits a continuous phase transition~\cite{Aizenman2015Random}.
	
Recently, based on theoretical intuition and numerical results,	the authors in Refs.~\cite{Fang2022Geometric, fk_cg_hd} argued that the FK Ising model has two upper critical dimensions $\dc$ = 4 and $\dpp$ = 6, depending on which quantities to be considered. They further found, as long as $d > 4$, the FK Ising model exhibits two-length-scale behavior, two configuration sectors, and two scaling windows. The scaling behaviors are simultaneously governed by the CG asymptotics and the GFP asymptotics for $4<d<6$, 
but with the GFP asymptotics replaced by the high-dimensional (high-d) percolation behavior for $d\ge 6$,
as summarized in Table~\ref{tab:FK_two_upper_critical_dimension}. This finding significantly enriches the understanding to the Ising model from a geometric perspective. Thus, one natural question is whether one can observe two upper critical dimensions in the loop Ising model, and whether the loop Ising model exhibits similar rich phenomena as the FK Ising model.

Before discussing the loop Ising model on tori, we first review known results on the CG, since it is believed that the scaling behaviors on high-d tori follow the CG asymptotics. In Ref.~\cite{loop_cg}, it was numerically found that, in contrast to the FK Ising model, the novel properties like the two-length-scale behavior, two configuration sectors, and two scaling windows cannot be explicitly observed in the loop Ising model on the CG. The sizes of the first- and second-largest loop clusters scale similarly
as $F_1, F_2 \sim V^{1/2}$ at the critical point $K_c = 1/V$. The cluster number density $n(s,V)$ was observed to behave as
	\begin{equation} 
		n(s,V) \asymp \frac{1}{2} V^{-1} \; s^{-1} \tilde{n} (s/\sqrt{V}), 
		\label{eq:ns_CG}
	\end{equation}
	where the scaling function $\tilde{n}(x \to 0)=1 $. 
 It follows that the total number of loop clusters $N_k = V \int n(s,V) ds \asymp \frac{1}{4} \ln V$, which is confirmed numerically. Further, they found loop configurations are asymptotically empty (bond density tends to 0 as $V\rightarrow \infty$), and after the LC transformation, many large loop clusters are merged together to form the largest FK cluster. Combining with the fact $\tanh{K_c} \sim 1/V$, which is the critical point of bond percolation on the CG, thus one can see that other FK clusters are basically generated by placing bonds onto an almost empty graph, which is a critical percolation process. This explains perfectly the percolation effects observed in the FK Ising model~\cite{Fang2021Percolation}.

	In this work, we employ the lifted worm algorithm~\cite{lifted_worm} to simulate the loop Ising model on high-d tori from $d=5$ to 7.  
 We find there are also two-length-scale behavior, two configuration sectors, and two scaling windows for the loop Ising model, while it only has one upper critical dimension $d_c=4$. The main results are summarized in Table~\ref{tab:FK_two_upper_critical_dimension}.  \par

 In Ref.~\cite{fk_cg_hd}, it was observed that, for the 7D FK Ising model, some quantities suffer unusually strong finite-size corrections at the estimated critical point, and this may be attributed to the insufficient precision of the estimate of the critical point. Thus, we first simulate the 7D loop Ising model and obtain a more precise estimate $\Kc = 0.077\,708\,91(4)$ through a systematic FSS analysis, which improves over the previous best estimate $0.077\,708\,6(8)$  by 20 times. 
	
At criticality, we first study the sizes of the largest- and second-largest loop clusters, $F_1$ and $F_2$. The finite-size fractal dimensions $(d_\textsc{l1}, d_\textsc{l2})$ and the thermodynamic fractal dimensions ($d_\textsc{f1}$, $d_\textsc{f2}$) are defined via $F_1 \sim L^{d_\textsc{l1}}\sim R_1^{d_\textsc{f1}}$ and $F_2 \sim L^{d_\textsc{l2}}\sim R_2^{d_\textsc{f2}}$, where $R_1$ and $R_2$ are the \emph{unwrapped} radii of gyration for the two clusters. We numerically find that $\dLone = \dLtwo = d/2$, following the CG asymptotics $V^{1/2}$ by matching the volume $V=L^{d}$, and $\dFone = \dFtwo = 2 = y_t$, following the GFP asymptotics. Thus, in contrast to the FK Ising model, the two largest clusters in the loop Ising model exhibit the same scaling behavior. For other loop clusters, our data suggest that the scaling $s \sim R^2$ also holds. In addition, it follows directly from the scaling of fractal dimensions that $R_{1,2} \sim L^{d/4}$, which means that loop clusters wind around the torus extensively for $d>4$. This is also different from the FK Ising model, in which the extensive winding does not happen until $d > 6$. Namely, in terms of winding, $6$ is a special dimension for the FK Ising model but not for the loop Ising model.

We then investigate the cluster-number density $n(s, L)$ of the loop Ising model, and our data suggest that for $d > 4$, it should be written as
	\begin{equation}
		n(s,L) \sim n_0 s^{-(1+d/2)} \tilde{n}_0(s/L^2) + n_1 s^{-1}L^{-d} \tilde{n}_1(s/L^{d/2}).  
		\label{eq:ns_tori}
	\end{equation}
Here, $\tilde{n}_0(x)$ and $\tilde{n}_1(x)$ are the scaling functions, and $n_0, n_1$ are two constants.
Clearly, two length scales can be observed from $n(s, L)$. For loop clusters with size $s\leq O(L^2)$, $n(s, L) \sim s^{-\tau}$ with the Fisher exponent $\tau = 1 + d/d_\textsc{f}$ and $d_\textsc{f} = 2$ from the GFP asymptotics, while for large loop clusters with size $s\geq O(L^2)$, $n(s, L)$ follows the CG behavior shown in Eq.~\eqref{eq:ns_CG}. Since the scaling $s \sim R^2$ holds for all loop clusters, the two length scales can also be interpreted as whether the radius of a cluster exceeds the system size $L$ (spanning or not). The number of spanning clusters $N_s$ can be obtained as $N_s \sim L^d \int_{L^2} n(s,L) {\rm d}s \sim \ln L$.

The FK Ising model on tori above 4D was found to have a special sector in the configuration space, in which quantities exhibit the GFP behavior for $4 < d < 6$ and the high-d percolation behavior for $d \geq 6$~\cite{Fang2021Percolation,fk_cg_hd}. For the loop Ising model, our data suggest that there also exists a special configuration sector which consists of loop configurations with the size of the largest loop cluster less than $L^2$. This sector exhibits the GFP behavior, and vanishes asymptotically with the rate $L^{1-d/4}$ for all $d > 4$. 
Moreover, we argue that, similar to the spin representation, the Ising model in the loop representation also has two scaling windows near the critical point; the narrow one is the CG-Ising window with width $O(L^{-d/2})$ and the wide one is the GFP window with width $O(L^{-2})$.
In the CG-Ising scaling windows, all quantities have the same scaling behaviors as it at the critical point. While in the GFP scaling window, the CG asymptotics of quantities are absent. For example, the average value of the first- and second-largest clusters $F_1 \sim F_2 \sim R_{1,2}^2 \sim L^2$, and the cluster number density $n(s,L)$ only  has the first term in the RHS of Eq.~\eqref{eq:ns_tori}.

From above, although there are also two-length-scale behavior, two configuration sectors and two scaling windows in the loop Ising model on high-d tori, unlike the FK Ising model, $d_p = 6$ is not a special dimension for the loop Ising model. Specifically, the loop Ising model has only one upper critical dimension $d_c = 4$, same as the spin Ising model. However, since the FK and loop Ising models are connected via the LC joint model, one would wonder why the FK Ising model has two upper critical dimensions $d_c = 4$ and $d_p = 6$, and especially, why there emerges percolation behavior above 6D. From the LC algorithm, an FK bond configuration can be generated by placing bonds with probability $\tanh K$ on the empty edges of a loop configuration. Our data show that, after placing bonds, the loop clusters with sizes larger than $O(L^2)$ are merged together and form the largest FK cluster.
This explains why in the FK Ising model, the largest cluster is much larger than other clusters. 
For dimensions $d > 6$, after large loop clusters are connected after placing bonds, other loop clusters are relatively small, and thus the bond placing process is like a percolation process and generates a large amount of FK clusters with no loop clusters involved in, which explains why above 6D the FK clusters, except the largest one, exhibit high-d percolation behavior.
	
	
	The remainder of this paper is organized as follows. In Sec.~\ref{sec:simulation}, the simulation details and sampled quantities are described. Section ~\ref{sec:loop} presents our numerical results.
	Finally, we sum up these results and provide a unified understanding to the scaling behaviors of the high-d Ising model in three representations in Sec.~\ref{sec:discussion}.

	\section{Simulation and Observable}
	\label{sec:simulation}
	
	\subsection{Algorithm}
	
In this section, we introduce the algorithms used in this work. We first employ the lifted worm algorithm~\cite{lifted_worm} to generate the loop configurations. Given a loop configuration, we use the LC algorithm~\cite{Zhang2020Loop} to generate a FK bond configuration, i.e., independently place bonds on the empty edges with probability $\tanh K$.
	
The worm algorithm is a type of Metropolis algorithm, which can efficiently update loop configurations via local moves. Given a loop configuration, a worm is located at a uniformly chosen vertex. Then, the worm tail is fixed at the vertex and the worm head performs a random walk on the lattice. At each step, the worm head proposes to walk through a uniformly chosen adjacent edge. If the edge is empty (not occupied by a bond), the proposal is accepted with probability $\tanh{K}$, and the edge becomes occupied after the worm head walks past. If the edge is occupied, the proposal is accepted with probability $1$ and the edge becomes unoccupied after the worm head walks through. It can be seen that if the head and tail are not on the same vertex, the bond configuration is not a loop configuration since the worm head and tail have odd degrees. Once the worm head hits the tail, a loop configuration is obtained.

In Ref.~\cite{lifted_worm}, the authors proposed a \emph{lifted} worm algorithm which is an irreversible Markov process. The idea of the lifted worm algorithm is to introduce an auxiliary variable $\lambda\in \{+,-\}$. When $\lambda = + (-)$, only the proposal of adding (removing) a bond is allowed. The parameter $\lambda$ flips whenever the proposal is rejected. It was shown in Ref.~\cite{lifted_worm} that the lifted worm algorithm is generally more efficient than the standard worm algorithm, especially in high dimensions and on the complete graph. According to Ref.~\cite{lifted_worm}, as the spatial dimension increases, the improving factor of the irreversible algorithm over the standard worm algorithm increases.  The integrated correlation time of the lifted worm algorithm (in unit of sweep)  is already smaller than one sweep, and exhibits critical speeding-up on the complete graph--i.e., $\tau_{\rm int} \sim 1/\sqrt{V}$ with $V$ the total number of vertices.

	\subsection{Sample quantities}
	
	\begin{table}[b]
		\centering
       \renewcommand{\arraystretch}{1.3}
		\begin{tabular}{|l|p{2.5cm}|p{1.5cm} |p{1.5cm}|}
			\hline 
			$d$ &   $K_c$ &$V_{\rm max}$  &$N_{\rm sam}$  \\
			\hline 
			5& 0.113\,915\,0(4)  \cite{Blote1997Universality} & $32^5 \approx 10^7$   & $ 6\times 10^5$ \\
			6& 0.092\,298\,2(3) \cite{Lundow2015Discontinuity} & $20^6 \approx 10^8$   & $3 \times 10^5$ \\
			7&  0.077\,708\,91(4) (this work) & $18^7 \approx 10^9$   & $3 \times 10^5$ \\
			\hline 
		\end{tabular}
		\caption{The critical points $K_c$ and the largest simulated system volume $V_{\rm max}$ for $d=5$, 6, 7. For each system, no less than $N_{\rm sam}$ independent samples are generated.}
		\label{tab:simulations}
	\end{table}
	We simulate the Ising model on high-d tori with $d = 5, 6, 7$. The critical points $\Kc$, the largest system volume $V_{\rm max}$, and the number of independent samples $N_{\rm sam}$  are summarized in Table~\ref{tab:simulations}. For each system, the following observables are sampled:     
	
	\begin{enumerate}[label=(\alph*)]
		\item The indicator $\mathcal{P}_{m}=1$ when the worm head hits the worm tail, i.e., a loop configuration is obtained, otherwise $\mathcal{P}_{m}=0$;
		
		\item The sizes of the first- and second-largest loop clusters denoted as $\mathcal{F}_1$ and $\mathcal{F}_2$;
		\item The number of loop clusters $\mathcal{N}(s)$ with size $s$, defined as the number of loop clusters with size in $[s, s + \Delta s)$ with an appropriately chosen interval size $\Delta s$;
		\item For a loop cluster $F$, its \emph{unwrapped} radius of gyration $\mathcal{R}(F)$ is defined as 
		$$\mathcal{R}(F) = \sqrt{\sum_{u\in F} \frac{({\bf x}_u - \bar{{\bf x}})^2}{|F|} },$$
		where $\bar{{\bf x}} = \sum_{u\in F} {\bf x}_u / |F|$. Here ${\bf x}_u \in \mathbb{Z}^d$ is defined algorithmically as follows. First, choose the vertex, say, $o$, in $F$ with the smallest vertex label according to some fixed but arbitrary labeling. Set ${\bf x}_{o} = 0$. Start from the vertex $o$, and search through the cluster $F$ using breadth-first growth. Iteratively set ${\bf x}_{v} = {\bf x}_{u} + {\bf e}_{i}(-{\bf e}_{i})$ if the vertex $v$ is traversed from $u$ along (against) the  $i$th direction, where ${\bf e}_i$ is the unit vector in the $i$th direction. The radii of the largest and the second-largest clusters are denoted as $\mathcal{R}_1$ and $\mathcal{R}_2$; 
		\item The largest unwrapped extension for each cluster in the first coordinate direction $\mathcal{U} = {\rm max}_{u,v \in F} ({\bf x}_u - {\bf x}_v)_{1}$;  
		\item The average radius of gyration for loop clusters with size in $[s,s+\Delta s)$
		$$\mathcal{R}(s) = \frac{\sum_{F:|F|\in [s+\Delta s)} \mathcal{R}(F)}{\mathcal{N}(s)};$$
		\item The number of spanning loop clusters $\scrN_s$. A loop cluster $F$ is spanning if $\scrU(F) \geq L$. We also sample $\scrN_1$ and $\scrN_2$, which are the number of loop clusters with size larger than $L^2$ and $2L^2$, respectively. 
		\item The total mass of large loop clusters $\scrS_{L^2} = \sum_{F: |F| \geq L^2} |F|$;
		\item The indicator  $\mathcal{P}=1$ when $\mathcal{F}_1\leq L^{2}$ is satisfied, otherwise $\mathcal{P}=0$. 
	\end{enumerate}

 	\begin{figure}[b]
		\centering
		\includegraphics[scale=0.7]{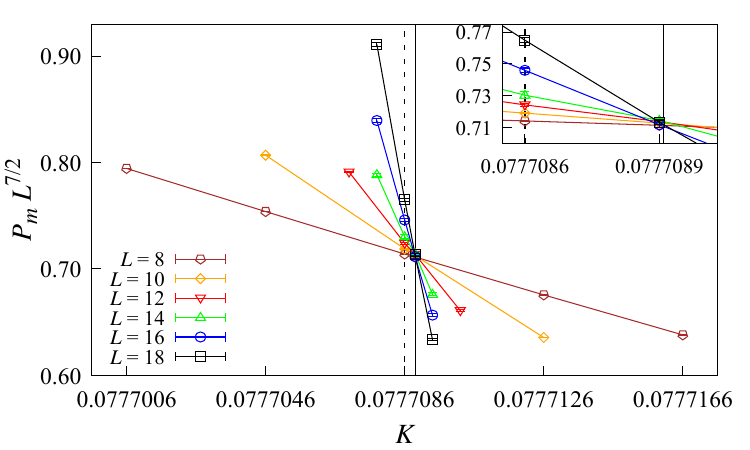}	
		\caption{Plot of the rescaled returning probability $P_m L^{7/2}$ of the 7D Ising model near the critical point and for various system sizes. The solid vertical line indicates the central value of the new estimate of $K_c = 0.077\,708\,91(4)$, while the dashed vertical line is the central value of the previous estimate of $K_c = 0.077\,708\,6(8)$. The inset clearly shows that our new estimate has higher precision.}
		\label{fig:Kc}
	\end{figure}
 
	From these observables, we can measure the following ensemble averages $\left\langle \cdot \right\rangle$:

	\begin{enumerate}[label=(\alph*)]
		\item The returning probability $P_m=\left\langle \mathcal{P}_m\right\rangle$;
		\item The mean sizes of the largest and the second-largest loop clusters $F_1=\left \langle\mathcal{F}_1\right \rangle $, $F_2=\left \langle \mathcal{F}_2\right \rangle $;
		\item The radius of gyration $R(s)=\left\langle\mathcal{R}(s)\right\rangle$  with given cluster size $s$. The mean radius of gyration of the largest and the second-largest loop clusters $R_1=\left \langle \mathcal{R}_1\right \rangle $, $R_2=\left \langle \mathcal{R}_2\right \rangle $;
		\item The mean number of spanning loop clusters $N_s = \langle \scrN_s \rangle$, and $N_m = \langle \scrN_m \rangle$ with $m \in \{1,2\}$; 
		\item The cluster-number density $n(s,L)=\frac{1}{L^d \Delta s}\left\langle\mathcal{N}(s)\right\rangle$;
		\item The probability of the configurations satisfying $\mathcal{F}_1\leq L^2$ denoted as $P = \left \langle \mathcal{P}\right \rangle$.
	\end{enumerate}
	
In addition to sample quantities in the loop configurations, we also sample the FK clusters. Denote $\mathcal{S}_{C_1}$ the total size of large loops (with size $\geq L^2$) that enters the largest FK cluster $C_1$, i.e., $\mathcal{S}_{C_1} = \sum_{F: F \subset \mathcal{C}_1, |F| \geq L^2} |F|$. We are interested in $n_f := \frac{\langle \scrS_{C_1} \rangle }{\langle \scrS_{L^2}\rangle}$, which is the fraction of vertices in the large loop clusters merged into the largest FK cluster, after the LC transformation. 

	\section{Results}
	\label{sec:loop}
	\subsection{Estimate of the critical point for $d=7$}
	
We first estimate the critical point $\Kc$ for the  7D Ising model by studying the FSS behavior of the returning probability $P_m$, which is expected to suffer from weaker finite-size corrections. For the worm algorithm, $P_m$ is identical to the reciprocal of the susceptibility, i.e., $P_m = 1/\chi$~\cite{Deng2007Dynamic}. Since $\chi \sim L^{d/2}$~\cite{Brezin1985Finite,Zhou2018Randomlength} above 4D, it follows that $P_m \sim L^{-d/2}$.

In Fig.~\ref{fig:Kc}, we plot the rescaled returning probability $P_m L^{7/2}$ versus the coupling strength $K$ with various system sizes. 
We find data from each studied system intersects around $K =0.077\, 708\,9$, which slightly deviates from the previous estimate $K_c =0.077\, 708\,6$. The inset zooms into this region and clearly shows this deviation.

	\begin{figure}[b]
		\centering
		\includegraphics[scale=0.7]{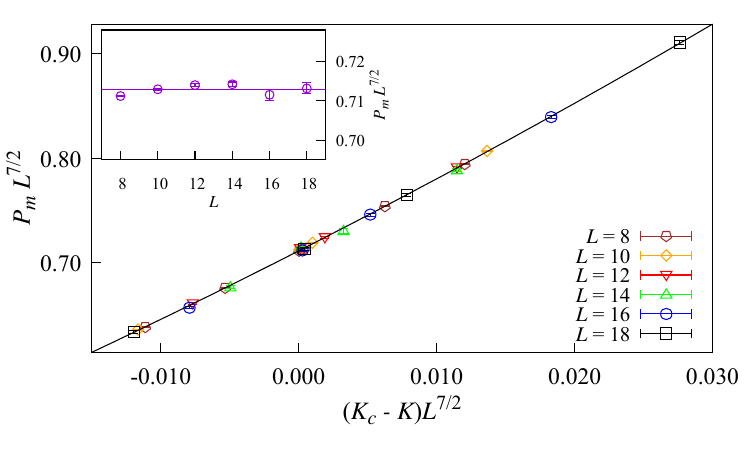}	
		\caption{Plot of $P_m L^{7/2}$ versus $(K_c - K)L^{7/2}$ to show the scaling function. The black curve corresponds to our preferred fit of the $P_m$ data to the ansatz Eq.~(\ref{eqs:fit}). The inset shows, at the estimated $K_c$, the data of $P_m L^{7/2}$ converge to a constant, which supports the accurancy of our estimate of $K_c$. }
		\label{fig:Kc_check}
	\end{figure}
 
To systematically estimate the critical point $K_c$, we perform least-squares fits of the MC data for the returning probability $P_m$ via the ansatz
	\begin{align}
		P_m L^{7/2} &= \sum_{k=0}^m q_k \left[ (K_c -K) L^{y_t}\right]^{k} + b_1 L^{y_1}   \nonumber \\
		&+ b_2 L^{y_2} + c_1(K_c -K) L^{y_t+y_1},
		\label{eqs:fit}
	\end{align}
	where $m$ is the highest order we keep in the fitting ansatz, $y_t$ is the thermal scaling exponent, and $y_2<y_1<0$ are finite-size correction exponents. The last term accounts for the crossing effect between the corrections and scaling variables. 
	
	\begin{table}[t]
		\centering
		\begin{tabular}{lclllcc}
			\hline \hline 
			$L_{\rm min}$  	&$y_t$ 	&~~~~~~~~$K_c$ 	&~~~~~$q_0$ 	&~~~$q_1$ 	&$q_2$	& $\chi^2/{\rm DF}$ 	\\
			\hline 
			10    &3.51(1)   	&0.077\,708\,906(5)	&0.712\,1(3) 	&6.6(2)    	&16(3)       	&19/15\\ 
			12    &3.51(3)   	&0.077\,708\,907(7)	&0.711\,8(5) 	&6.6(5)    	&18(5)        	&17/11\\ 
			14    &3.51(7)   	&0.077\,708\,92(1)	&0.710(1)  	    &6(1)      	&17(8)       	&11/7\\  
			10    &7/2      	&0.077\,708\,906(4)	&0.712\,1(3) 	&6.74(2)   	&17(2)        	&19/16\\ 
			12    &7/2      	&0.077\,708\,907(7)	&0.711\,8(5) 	&6.73(4)   	&19(3)      	&17/12\\ 
			14    &7/2      	&0.077\,708\,92(1)	&0.710(1)  	    &6.73(6)   	&18(3)       	&11/8\\  
			\hline
			\hline 
		\end{tabular} 
		\caption{Fitting results for the returning probability $P_m$ using the ansatz Eq.~(\ref{eqs:fit}) with $m=2$, and $b_1,b_2,c_1 = 0$.}
		\label{tab:Kc_fit} 
	\end{table}

	As a precaution against correction-to-scaling terms that we missed including in the fitting ansatz, we impose a lower cutoff $L \ge L_{\rm min}$ on the data points admitted in the fit and systematically study the effect on the residuals $\chi^2$ value by increasing $L_{\rm min}$. In general, the preferred fit for any given ansatz corresponds to the smallest $L_{\rm min}$ for which the goodness of the fit is reasonable and for which subsequent increases in $L_{\rm min}$ do not cause the $\chi^2$ value to drop by vastly more than one unit per degree of freedom. In practice, by “reasonable” we mean that $\chi^2/\rm{DF} \approx 1$, where DF is the number of degrees of freedom. The systematic error is estimated by comparing estimates from various sensible fitting ansatz.  

Firstly, we try to fit by setting $m=2$ and leaving all other parameters free, but it gives unstable results. Then, by fixing $b_1$, $b_2$, $c_1 = 0$, the fitting shows that $\chi^2/{\rm DF} \approx 1$ when $L_{\rm min} = 10$ and gives $K_c = 0.077\,708\,906(5)$ and $y_t = 3.51(1)$, consistent with the expected value $7/2$. This implies that finite-size corrections for $P_m$ are indeed quite weak. Including higher order terms to Eq.~\eqref{eqs:fit} gives that the coefficients $q_k$ are consistent with zero when $k\ge3$. Thus, in the following, we fix $y_t = 7/2$ and $m = 2$. 
	
We then perform the fits by leaving $b_1$, $b_2$, $y_1$, $y_2$ free, and again it fails to produce stable fits. Fixing $y_1=-1$, $y_2=-2$ or $y_1=-2$, $y_2=-3$ produce consistent estimates of $K_c = 0.077\,708\,94(3)$ and $K_c = 0.077\,708\,94(2)$, respectively. In all scenarios above, including the crossing-effect term to the ansatz shows that $c_1$ is consistent with zero, and its effect on the estimates of other parameters is negligible. Fitting results without any correction terms are shown in Table~\ref{tab:Kc_fit}. By comparing estimates from various ansatz, we conclude that $K_c = 0.077\,708\,91(4)$. Figure~\ref{fig:Kc_check} shows the data of $P_mL^{7/2}$ versus the scaling variable $(K-K_c)L^{7/2}$, and all data collapse nicely onto the curve which corresponds to our preferred fitting to the ansatz Eq.~\eqref{eqs:fit}. Furthermore, the inset displays $P_m L^{7/2}$ versus $L$ at our estimated $K_c$, which converges to a constant as $L$ increases. This is consistent with the expected scaling $P_m \sim L^{-7/2}$ at the critical point. 
	
	\subsection{The fractal dimensions of loop clusters}
	In this section, we study the fractal dimensions of loop clusters for $d=5$, 6, 7. Inspired by the FK Ising model~\cite{fk_cg_hd}, we consider the finite-size fractal dimensions $(d_\textsc{l1}, d_\textsc{l2})$ and the thermodynamic fractal dimensions  $(d_\textsc{f1}, d_\textsc{f2})$ for the largest and the second-largest loop clusters, which are defined as $ F_1 \sim L^{d_\textsc{l1}} \sim R_1^{d_\textsc{f1}}$ and $ F_2 \sim L^{d_\textsc{l2}} \sim R_2^{d_\textsc{f2}}$ with the loop cluster sizes $F_1$, $F_2$ and their \emph{unwrapped } radii $R_1$, $R_2$. 

	\begin{figure}[t]
		\centering
		\includegraphics[scale=0.65]{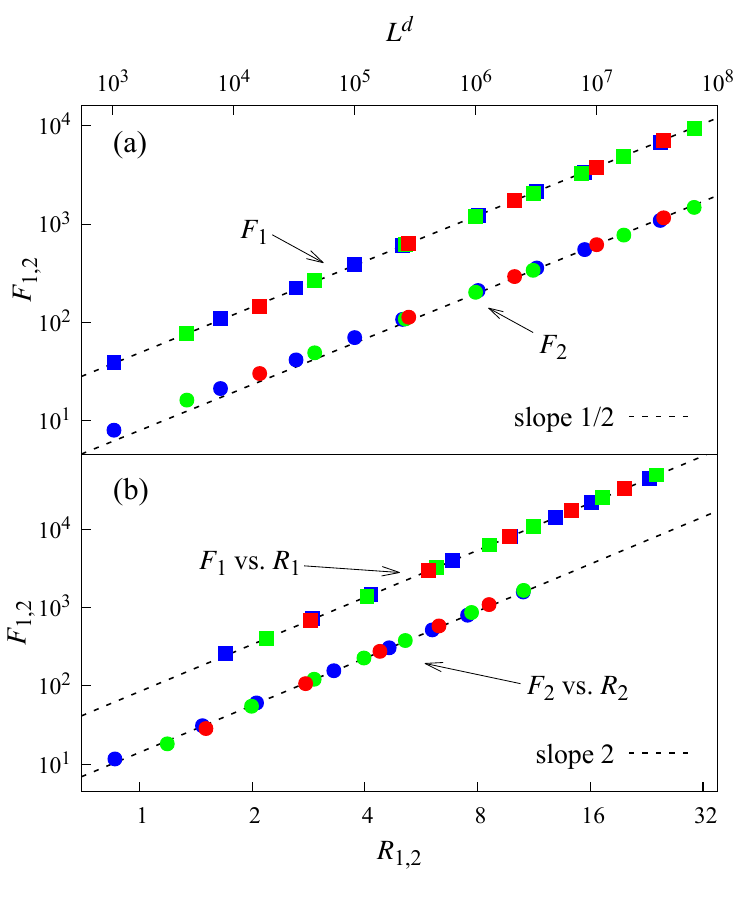}	
        \caption{The log-log plot of the largest loop cluster $F_1$ and second-largest loop cluster $F_2$ versus (a) system volume $L^{d}$ and (b) their radii $R_{1,2}$ for $d=5$ (blue), $d=6$ (green) and $d=7$ (red). It implies  the finite-size  fractal dimensions $d_\textsc{l1} = d_\textsc{l2}= d/2$  and thermodynamic fractal dimensions $d_\textsc{f1} = d_\textsc{f2}= 2$.}
		\label{fig:fractal_dim}
	\end{figure}

	For the finite-size fractal dimensions, we first recall that the authors in Ref.~\cite{loop_cg} found, on the CG,  both the first- and second-largest loop clusters have the same scaling behavior $F_1, F_2 \sim V^{1/2}$.
	By matching $V=L^{d}$, we expect $F_1, F_2 \sim L^{d/2}$ on high-d tori. 
	In Fig.~\ref{fig:fractal_dim}(a), we plot loop clusters $F_1$ and $F_2$ versus the system volume $L^{d}$. In the log-log scale, both the data of $F_1$ and $F_2$ from various spatial dimensions collapse onto lines with slope $1/2$, which indicates $d_\textsc{l1}= d_\textsc{l2}=d/2$, following the CG asymptotic.

As for the thermodynamic fractal dimensions, we consider the unwrapped radii of the largest and the second-largest clusters $R_1$ and $R_2$.  In Fig.~\ref{fig:fractal_dim}(b), we plot the loop clusters $F_1$ and $F_2$ versus their radii in the log-log scale. Data from various dimensions collapse well onto a straight line with slope $2$, which implies $d_\textsc{f1} = d_\textsc{f2}= 2$. We note that these two exponents are equal to the GFP exponent $y_t=2$, which can be understood as follows. In high dimensions, one can expect that large loop clusters are mostly self-avoiding polygons (or unicycles), as on the complete graph~\cite{loop_cg}. For self-avoiding polygons, which is in the same universality as the self-avoiding walk, it is known that for $d > 4$, the size scales as the square of the radius of gyration~\cite{Madras2013Self}. Thus, the same scaling behavior is expected for the loops in the loop Ising model.


		\begin{figure}[t]
		\centering
		\includegraphics[scale=0.65]{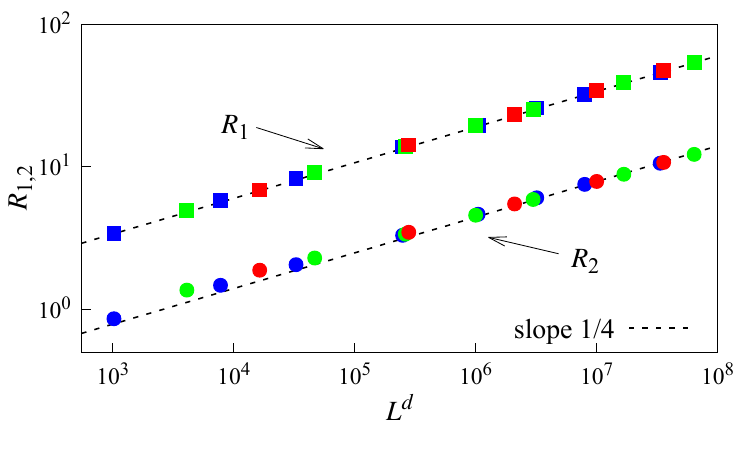}	
		\caption{The log-log plot of the radii of the largest and the second-largest clusters $R_1$ and $R_2$ versus $L^d$ for $d=5$ (blue), $d=6$ (green) and $d=7$ (red). It implies the scaling behavior $R_1, R_2 \sim L^{d/4}$.}
		\label{fig:R_L}
	\end{figure}

 	\begin{figure}[b]
		\centering
		\includegraphics[scale=0.65]{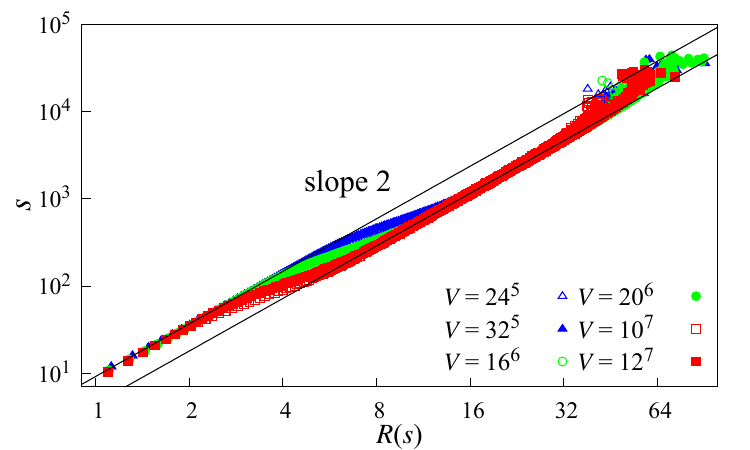}	
		\caption{The log-log plot of the cluster size $s$ versus its radii $R(s)$ for all loop clusters and $d\geq5$. It implies that the scaling behavior $s\sim R^2$ holds for small and large loop clusters, with a crossover in between.}
		\label{fig:sR}
	\end{figure}

	\begin{figure*}
		\centering
		\includegraphics[scale=0.55]{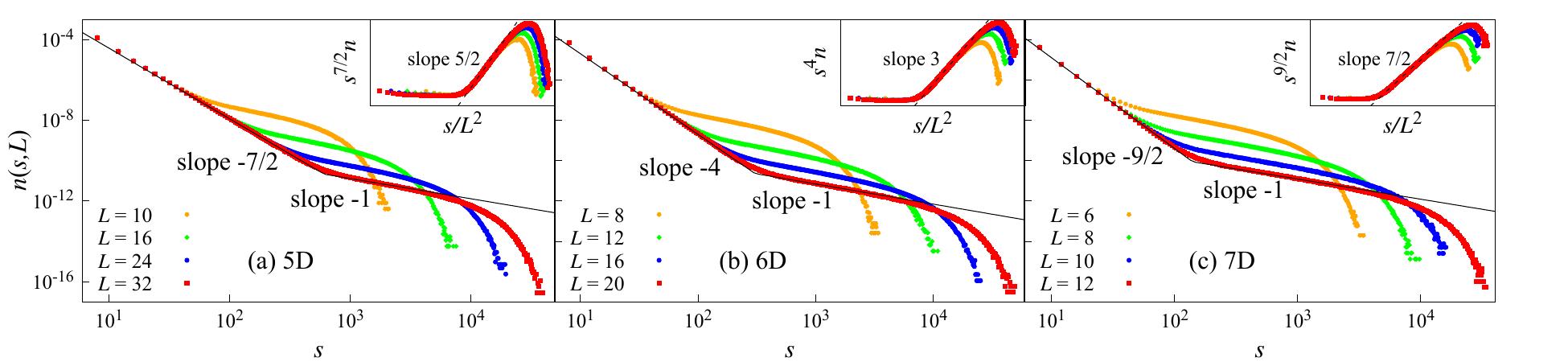}	
  \caption{The log-log plots of the cluster number density $n(s,L)$ for (a) $d=5$,  (b) $d=6$ and (c) $d=7$. The inset in each subfigure displays the plot of $n(s,L)s^{1+d/2}$ versus $s/L^2$ for each studied dimension. It indicates that there are two length scales in $n(s, L)$. The scale corresponding to small loop clusters ($s < O(L^2)$) follows the GFP asymptotics.}
		\label{fig:HD_ns}
	\end{figure*}
 
Since $F_1, F_2 \sim L^{d/2}, F_1\sim R_1^2$ and $F_2\sim R_2^2$, we expect $R_1\sim R_2\sim L^{d/4}$, which is larger than the system size $L$ for $d>d_c=4$. In Fig.~\ref{fig:R_L}, the plot of cluster sizes $F_1$, $F_2$ versus their radii $R_1$, $R_2$ collapses well onto straight lines with slope consistent with $1/4$. This scaling behavior indicates that large loop clusters wind around the boundary many times for $d>4$. This is different from the observation in the FK Ising model, in which $R_1\sim L$ for $d\leq 6$ and $R_1\sim L^{d/6}$ for $d>6$~\cite{fk_cg_hd}.

We also investigate the thermodynamic fractal dimensions for all loop clusters and plot their sizes $s$ versus their radii $R$ in Fig.~\ref{fig:sR}. 
It can be seen that the scaling $s \sim R^2$ holds for both small and large loop clusters, with a crossover happening in between. We argue that the fractal dimension of all clusters is $d_\textsc{f} =2$, and the scaling behavior of these medium-size clusters in the crossover region are due to the boundary effect.    
Namely, this region is the crossover between the CG asymptotics for large clusters and the GFP asymptotics for small clusters, and loops of size $O(L^2)$ or smaller start to merge together and form large loop clusters. Nevertheless, the power-law dependence of loop-cluster size $s$ on gyration radius $R$ still satisfies $s \sim R^2$.

Therefore, for $d>4$, the finite-size fractal dimensions of the first- and the second-largest loop cluster are consistent with $d/2$, following the CG asymptotics, and the thermodynamic fractal dimensions of all loop clusters are consistent with 2, following the GFP asymptotics. 
From the perspective of fractal dimensions, $d_p = 6$ is not a special dimension for the loop Ising model.
	

	\subsection{The cluster-number density}
In this section, we study the cluster-number density $n(s,L)$. In Fig.~\ref{fig:HD_ns}, we plot $n(s,L)$ versus cluster size $s$ in log-log scale and find it exhibits two scaling behaviors for each studied spatial dimension, which is similar to the bridge-free configurations of the high-d percolation model~\cite{perco_7d}.
For small $s$, $n(s, L)$ shows a power-law decay with exponents consistent with $-7/2$ at 5D, $-4$ at 6D and $-9/2$ at 7D. We note that these power-law exponents are consistent with $1+d/2$ for $d=5, 6$ and $7$. For large $s$, at each dimension, the data of $n(s, L)$ fails to collapse for various systems. For a given dimension and system size, $n(s, L)$ still exhibits the power-law behavior but with a constant exponent $-1$.

How to understand the two scaling behaviors of the cluster number density $n(s,L)$? Generally, for $n(s,L)$, it is believed that it follows 
\begin{equation}
n(s,L) = n_0 s^{-\tau} \tilde{n}(s/L^{d_f})\quad \left[\Tilde{n}(x\rightarrow0)=1\right]\;,
\label{Eq:cluster-numbero-density-general}
\end{equation}
where $n_0$ is a positive constant, $\tau$ is the Fisher exponent, $\tilde{n}(\cdot)$ is the scaling function, and $d_f$ is the fractal dimension of the largest cluster. Usually, the Fisher exponent obeys the hyperscaling relation 
	\begin{equation}
		\tau = 1+ d/d_f. 
	\end{equation}
Equation~\eqref{Eq:cluster-numbero-density-general} has been observed for the loop Ising model in two and three dimensions~\cite{Liu2011Worm,Liu2012loop}.      
For $d>4$, we find that if $d_f= 2$, taking the GFP prediction, then it follows that $\tau = 1+d/2$, consistent with the small $s$ behavior in Fig.~\ref{fig:HD_ns}. The power-law exponent governing the scaling of $n(s, L)$ for large $s$ is $-1$, which is consistent with the CG case~\cite{loop_cg}. Thus, we conjecture the scaling behavior $n(s, L)$ follows Eq.~\eqref{eq:ns_tori}. Namely, the scaling behavior of $n(s,L)$ is simultaneously governed by the GFP prediction and the CG asymptotics; the former controls the power-law decay of small loop clusters while the latter controls the power-law decay of large loop clusters. It follows from Eq.~\eqref{eq:ns_tori} that the crossover happens at $s =O(L^2)$. Thus, although $n(s, L)$ exhibits the two-length-scale behavior, it suggests for the loop Ising model only $d_c = 4$ is the upper critical dimension.

\begin{figure}[h!]
		\centering
		\includegraphics[scale=0.65]{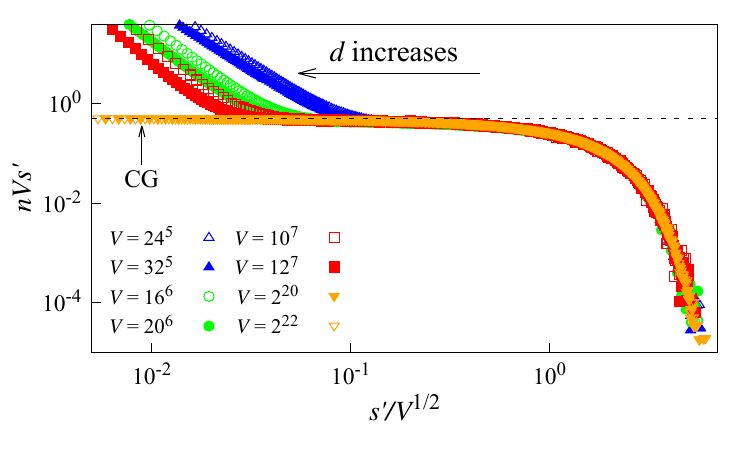}	
		\caption{The log-log plot of rescaled cluster number density $n(s,L)V s'$ versus $s'/V^{1/2}$, where $s' = s/\alpha$ is rescaled cluster size with  $\alpha = 1.2, 1.1, 1.05, 1$ for $d = 5, 6, 7$ and the CG, respectively. The good data collapse indicates $n(s,V)$ obeys the CG asymptotics for large loop clusters.} 
		\label{fig:HD_ns_3}
	\end{figure}

To verify Eq.~\eqref{eq:ns_tori}, we plot $n(s, L) s^{1+d/2}$ versus $s/L^2$, as shown in the inset of Fig.~\ref{fig:HD_ns}. It follows from Eq.~\eqref{eq:ns_tori} that
\begin{equation}
    s^{1 + d/2} n(s, L) \sim n_0 \tilde{n}_0 \left(\frac{s}{L^2}\right) + n_1 \left(\frac{s}{L^2}\right)^{d/2} \tilde{n}_1\left(\frac{s}{L^{d/2}}\right),  \nonumber
\end{equation}
Thus, $s^{1 + d/2} n(s, L)$ equals the constant $n_0$ if $s < O(L^2)$ and increases as a power-law with exponent $d/2$ if $s > O(L^2)$. This is consistent with the data shown in the inset of Fig.~\ref{fig:HD_ns}. To clearly show that the large loop clusters follow the CG asymptotics, in Fig.~\ref{fig:HD_ns_3} we plot the data of $n(s,L)V s'$ versus $s'/ V^{1/2}$ for each dimension and also for the CG to compare with. Here $s' = s/\alpha$ with $\alpha$ depending on $d$ so that the data of each dimension can collapse together. As Fig.~\ref{fig:HD_ns_3} shows, the data of $n(s,L)$ on high-d tori collapse nicely onto the CG data when $s'/\sqrt{V}$ is large. The discrepancy in the small $s'/\sqrt{V}$ part is due to the existence of the Gaussian length scale, in which typical large loops have size of order $L^2$. We expect such a discrepancy vanishes with the rate $L^{2-d/2}$, decaying faster for larger $d$ as shown in the figure. This can also be seen from the Gaussian Fisher exponent $\tau = 1 + d/2$. As $d\rightarrow \infty$, $\tau$ tends to infinity, such that the Gaussian part vanishes to zero and the system completely follow the CG asymptotics.


	\begin{figure}[b]
		\centering
		\includegraphics[scale=0.65]{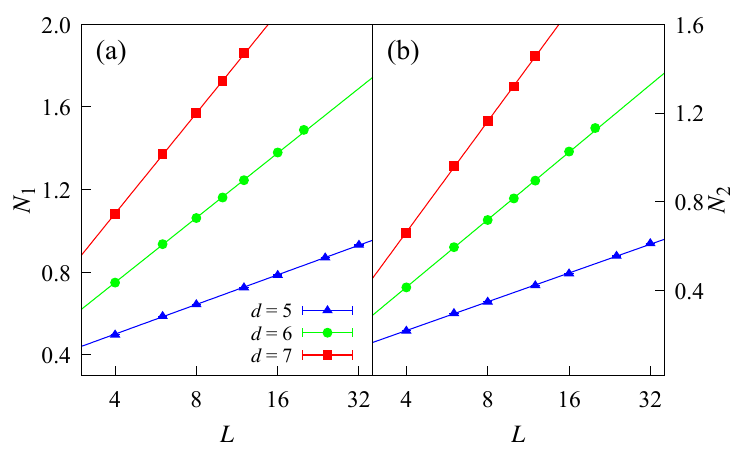}	
        \caption{The semi-log plot of  (a) $N_1$ and (b) $N_2$ versus  system size $L$ for $d\geq5$, where the cluster number $N_m (m=1,2)$ are the number of clusters that satisfies $s \geq m L^2$. The straight lines indicate both $N_1$ and $N_2$ diverge logarithmically.}
		\label{fig:N1_N2}
	\end{figure}

To further confirm our conjecture, we study $N_m$, the number of loop clusters with size $s > mL^2$. Since the large loop clusters follow the CG asymptotics, it follows that $N_m$ can be calculated as
\begin{align}
L^d \int_{m L^2}^{L^{d/2}} n(s,L) ds \sim \int_{m L^2}^{L^{d/2}} s^{-1} \tilde{n}(s/L^{d/2}) ds \sim \ln L. \nonumber
\end{align}
In simulations, we sample $N_1$ and $N_2$, and the data are plotted In Fig.~\ref{fig:N1_N2}. Clearly, it strongly suggests that both $N_1$ and $N_2$ scale as $\ln L$ for each studied dimension.

	\begin{figure}[h!]
		\centering
		\includegraphics[scale=0.65]{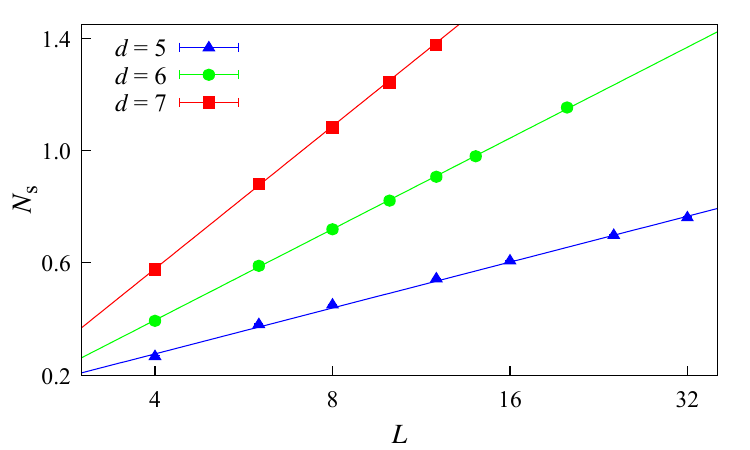}	
		\caption{The semi-log plot of the number of spanning cluster  $N_s$ versus the system size $L$ for $d \geq 5$. It indicates $N_s$ has the same scaling behavior with $N_1$ and $N_2$, i.e., $N_s, N_1, N_2 \sim \ln L$.}
		\label{fig:Ns}
	\end{figure}
	
Finally, we study the number of spanning clusters $N_s$ for $d \geq 5$. Recall that a cluster is called spanning if its unwrapped extension $\scrU \ge L$. It can be expected that the unwrapped extension and the unwrapped radius exhibit the same scaling behavior. From Fig.~\ref{fig:sR}, we know that a loop cluster is spanning if its size is larger than $O(L^2)$. Thus, it follows from $n(s, L)$ that $N_s \sim \ln L$, the same scaling as $N_1$ and $N_2$. In Fig.~\ref{fig:Ns}, the data of $N_s$ is plotted versus $L$ in the semi-log scale and clearly, it suggests that $N_s \sim \ln L$. Recall that for the FK Ising model, the number of spanning clusters is of constant order for $d < 6$ and diverges as $L^{d-6}$ for $d > 6$. But for the loop Ising model, $N_s$ diverges logarithmically for $d > 4$, again implying that $6$ is not a special dimension for the loop Ising model.

\subsection{Probability distribution of the largest loop cluster}

\begin{figure}[t]
\centering
\includegraphics[scale=0.65]{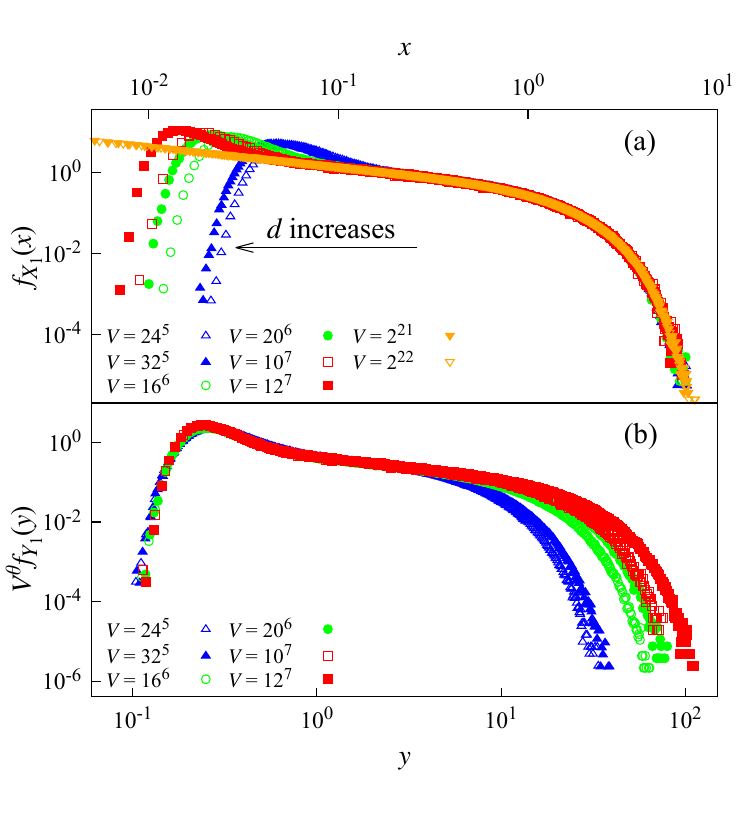}	
\caption{Plots of probability density functions of the largest loop cluster ${\mathcal{F}_1}$.  In subfigure (a),  the variable is defined as  $X_1 = {\mathcal{F}_1}/({aL^{d/2}})$, where the constant $a$ is chosen to be $1, 0.90,0.85, 0.8$ for $d = 5$ (blue), 6 (green), 7 (red) and CG (orange),  respectively.  In subfigure (b), the variable $Y_1 = {\mathcal{F}_1}/(b L^2)$ with $b =$1, 1.52 and 2.12   and $\theta =$  1/20, 1/12, 3/28 for $d = 5$, 6 and 7, respectively. It illustrates there is also a special vanishing sector in the configuration space and the probability distribution of the largest loop cluster obeys the CG asymptotics for $V \to \infty$. }
\label{fig:fX1}
\end{figure}

In this section, we study the probability density function of the largest loop cluster size $\scrF_1$ on high-d tori, which is denoted as $f_{\scrF_1}(s)$, and compare it with the CG case. Since $F_1\sim L^{d/2}$, we define $X_1 = {\scrF_1}/({aL^{d/2}})$ with a non-universal constant $a$ for each studied dimension $d$ and its probability density function as $f_{X_1}(x)$. It follows that
\begin{align}
 f_{\scrF_1}(s) ds = f_{X_1}(x) dx\;, \nonumber
\end{align}
where $dx = a^{-1} L^{-d/2} ds$ and thus $f_{X_1}(x) = aL^{d/2} f_{\scrF_1}(s)$. Figure~\ref{fig:fX1}(a) plots $f_{X_1}(x)$,  and it shows that when $x \gtrsim 0.2$, data from various spatial dimensions collapse well onto the CG data. Here the parameter $a$ is chosen to be 1, 0.90, 0.85 and  0.8, respectively for $d=5$, 6, 7 and the CG.	
	
However, as Fig.~\ref{fig:fX1}(a) shows, when $x$ is small, the data cannot collapse well and deviates from the CG data. But it seems as $d$ increases, the deviation becomes smaller. This is similar to the observation in the FK Ising model on high-d tori and CG~\cite{Fang2021Percolation,fk_cg_hd}, which is due to the existence of a special sector in the configuration space. Thus, we conjecture that there is also a special sector in the loop Ising model on high-d tori. From the behavior of $n(s, L)$ in Eq.~\eqref{eq:ns_tori}, we know small loop clusters with size $\leq O(L^2)$ obey the GFP asymptotics. Thus, we conjecture that the average size of loop clusters in the special sector is $O(L^2)$. We define $Y_1 = \mathcal{F}_1/(bL^2)$ with some $d$-dependent constant $b$.
Similarly, we have
\begin{align}
f_{F_1}(s)ds = f_{Y_1}(y) dy \;, \nonumber
\end{align}
where $dy = b^{-1} L^{-2}ds$ and thus $f_{Y_1}(y) = bL^2f_{F_1}(s)$. We then plot $f_{Y_1}(y)$ versus $y$, but the data show that it decays as a power-law as the system size increases. This implies that this special sector vanishes to 0 as $L \rightarrow \infty$. To find the power-law exponent, we assume that the probability $\PP(\scrF_1 \leq b L^2)$ on high-d tori has the same scaling as $\PP(\scrF_1 \leq V^{2/d})$ on the CG; the latter can be calculated explicitly as
\begin{align}
\int_1^{V^{\frac{2}{d}}} f_{F_1} (s) ds  \sim V^{\frac{1}{d}-\frac{1}{4}}.
\end{align}
where on the CG it was obtained in Ref.~\cite{loop_cg} that $f_{F_1}(s) \sim V^{-\frac{1}{4}} s^{-\frac{1}{2}} \Tilde{f}_{F_1}(s/V^{1/2})$ with $\Tilde{f}_{F_1}(\cdot)$ the scaling function. Thus, we conjecture the special sector in the loop Ising model vanishes with the rate $V^{1/d-1/4}$.

\begin{figure}[h!]
		\centering
		\includegraphics[scale=0.65]{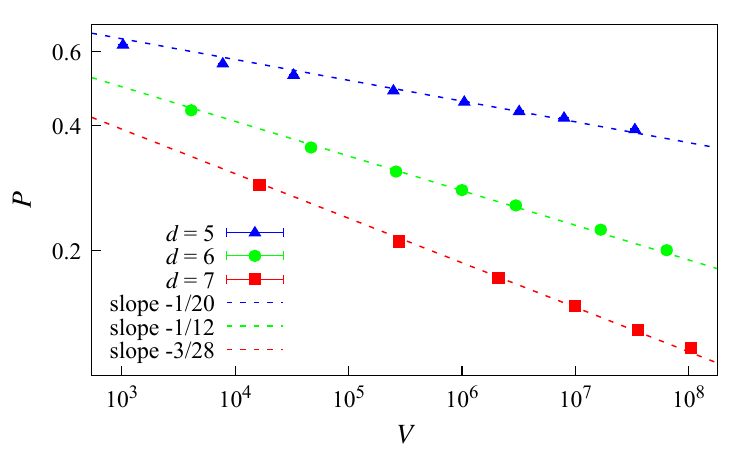}	
    \caption{Log-log plot of the probability of $P =\PP(\mathcal{F}_1 \leq L^2)$ versus the system volume $V$ for $d = 5, 6, 7$. It implies that the special sector of the loop Ising model vanishes as $V^{\frac{1}{d}-\frac{1}{4}}$. }
		\label{fig:two_sectors}
	\end{figure}

In Fig.~\ref{fig:fX1}(b), we plot $V^{1/d-1/4}f_{Y_1}(y)$ versus $y$. Indeed, the data from various spatial dimensions collapse well for small $y$. 
To verify our conjecture, we numerically study the probability $P = \PP(\scrF_1 \leq L^2)$ for $d = 5$, 6, 7. In Fig.~\ref{fig:two_sectors}, the data are plotted versus the system volume $V$ in the log-log scale, and the slopes are consistent with $-1/20$, $-1/12$ and $-3/28$ for $d=5, 6$ and $7$ respectively, which supports the conjecture $P \sim V^{1/d - 1/4}$.
In addition, one notes as $d \to \infty$, the vanishing rate $P \sim V^{-1/4}$, consistent with the observation on the probability of the empty graph in the CG loop Ising model~\cite{loop_cg}.

As shown in Table~\ref{tab:FK_two_upper_critical_dimension}, the vanishing sector in the FK Ising model decays as $L^{1-d/4}$ for $4 < d < 6$ but as $L^{-d/12}$ for $d > 6$. For the loop Ising model, our data show that the vanishing rate is $L^{1-d/4}$ for all $d > 4$. 
Note that the exponents $1-d/4 = y_h -y_h^*$ and $-d/12 = y_{h,\textsc{p}}^* -y_h^*$ with the GFP exponent $y_h = 1+d/2$, CG-Ising exponent $y_h^*=3d/4$, and CG-percolation exponent $y^*_{h,\textsc{p}}=2d/3$.
Again, it suggests that $6$ is a special dimension for the FK Ising model but not for the loop Ising model.

\subsection{Transformation from the loop representation to the FK representation }

\begin{figure}[t]
		\centering
		\includegraphics[scale=0.70]{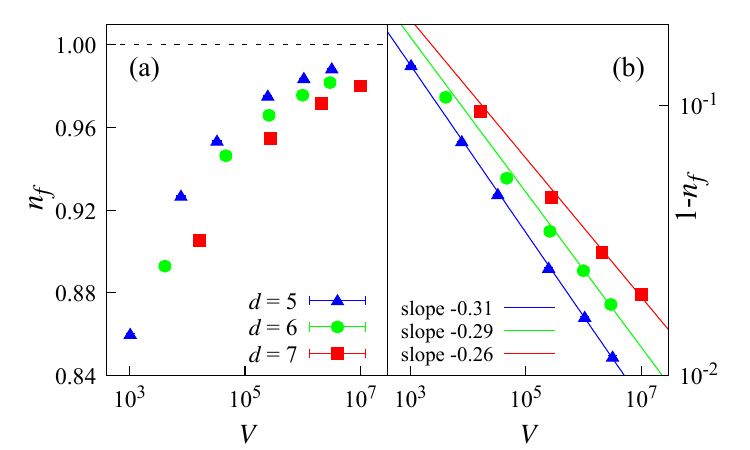}	
        \caption{(a) Plot of $n_f$, the percentage of large loop clusters (with size $\ge L^2$) in the largest FK cluster, as the system size $V$ increases. (b) Log-log plot of $1 - n_f$ versus $V$. It implies that, as $V \rightarrow \infty$, $1-n_f$ decays as a power-law with slopes approximately consistent with -0.31, -0.29 and -0.26, respectively. This suggests almost all the large loop clusters are merged into the largest FK cluster.}
		\label{fig:nf}
	\end{figure}

In this section, the connections between the loop representation and the FK representation in high dimensions are demonstrated. In the FK representation, the largest and second-largest clusters scale as $C_1 \sim L^{3d/4}$ and $C_2 \sim L^{1 + d/2}(\ln L)^{-1}$~\cite{fk_cg_hd}; both are much larger than the sizes of the two largest loop clusters in the loop Ising model which are $F_1, F_2 \sim L^{d/2}$. Since a typical FK bond configuration can be generated by placing bonds with probability $\tanh K$ onto a loop configuration, it is interesting to study how loop clusters are merged into FK clusters. Inspired by the two-length-scale behavior observed in $n(s,V)$, we conjecture that loop clusters with size $s\geq L^2$ are merged into the largest FK cluster after those extra bonds are placed. To check this numerically, we sample $n_f$, which is the percentage of loop clusters merged into the largest FK cluster, conditioned on that loop clusters have size larger than $L^2$. 
In Fig.~\ref{fig:nf}(a), we plot $n_f$ versus $V$ for $d\geq 5$ with semi-log plot, which shows that $n_f$ increases as $V$. 
To confirm $n_f$ converges to 1, we plot $1 - n_f$ versus $V$ in the log-log scale in Fig.~\ref{fig:nf}(b), which clearly shows that $1-n_f$ decays as a power-law and thus indeed $n_f$ converges to 1. It suggests that all loop clusters with size $\ge L^2$ are merged together to form the largest FK cluster asymptotically. We note that, the power-law exponents in Fig.~\ref{fig:nf}(b) are consistent with $-0.31$, $-0.29$, and $-0.26$, for $d=5$, 6, and 7, respectively. As $d\rightarrow \infty$, we expect it converges to the observed value $-0.225$ on the CG~\cite{loop_cg}.

In what follows, we term the largest FK cluster and the loop clusters with size of order $L^{d/2}$ as giant clusters, and the FK clusters with size of order $L^{1+d/2}$ and the loop clusters with size of order $L^2$ as medium-size clusters. All other clusters are called small-size clusters. We next discuss the connection between medium-size clusters in the loop and FK representations. The Fisher exponent governing the cluster-size distribution of the medium-size clusters is $\tau = 1 + d/2$ for the loop Ising model with $d > 4$, and for the FK Ising model $\tau = 1+\frac{d}{1+d/2}$ for $4<d<6$ and $\tau =5/2$ for $d \ge 6$~\cite{Fang2022Geometric, fk_cg_hd}. Denote $N_{\rm Loop}$ and $N_{\rm FK}$ the number of medium-size loop and FK clusters, respectively. It can be shown that both $N_{\rm Loop}$ and $N_{\rm FK}$ are $O(1)$ for $4<d<6$. Thus, we conjecture that on average each medium-size FK cluster contains $O(1)$ number of medium-size loop clusters. In other words, the medium-size FK clusters are mainly generated from the medium-size loop clusters, and thus both of them exhibit the GFP behavior. However, for $d > 6$, $N_{\rm Loop}$ is still $O(1)$ but $N_{\rm FK}$ diverges as $ L^{(d-6)/4}$. So on average, the medium-size FK clusters contain no medium-size loop clusters. Namely, for $d > 6$ almost all medium-size FK clusters are generated by the percolation-like process, and thus exhibit high-d percolation behavior. We expect this argument can also be used to explain the connection between smaller FK and loop clusters (smaller than medium-size clusters). Thus, we argue that for $d > 6$, all FK clusters except the largest cluster exhibit the same behavior as high-d percolation clusters, like the thermodynamic fractal dimension $d_\textsc{f}=4$ and the number of spanning clusters $N_s \sim L^{d-6}$. 

As Fig.~\ref{fig:HD_ns} shows, the loop Ising model has two length scales; giant loop clusters follow the CG asymptotics but other clusters follow the GFP asymptotics. After the LC transformation, as shown in Fig.~\ref{fig:nf}, all giant loops are merged together to form the largest FK cluster, and other loop clusters are transformed into other FK clusters. Thus, it is natural to expect there are two length scales in the FK Ising model~\cite{fk_cg_hd}.

We finally discuss the special configuration sectors. For the loop Ising model with $d > 4$, our data suggest that the special sector, consisting of loop configurations in which the largest loop cluster has size $O(L^2)$, accounts for a proportion $\sim L^{1-d/4}$ of the whole configuration space. By our conjecture, these medium-size loop clusters (size of order $L^2$) will become the medium-size FK clusters (size of order $L^{1+d/2}$), after the LC transformation. Since for $4 < d < 6$, all medium-size FK clusters are generated by medium-size loop clusters, it is natural to expect there exists a special configuration sector in the FK Ising model, which also vanishes with the rate $L^{1-d/4}$. This was numerically confirmed in Ref.~\cite{fk_cg_hd}, and in the special sector, all FK clusters were found to exhibit the GFP behavior. However, the scenario is more complicated for $d > 6$. On the CG (the $d\rightarrow \infty$ case), it was found that~\cite{loop_cg} the FK Ising model has a special configuration sector in which FK clusters exhibit the CG percolation clusters behavior, and this sector corresponds to the sector of loop configurations with the largest loop size of order $V^{1/3}$; both sectors asymptotically account for $V^{-1/12}$ of their own whole configuration space. Assume the CG results hold also on high dimensional torus. Then, since $L^{1-d/4} \ll L^{-d/12}$ when $d > 6$, it follows that loop configurations with the largest loop cluster of order $L^2$ are not enough to generate the special sector in the FK Ising model for $d>6$. Thus for $d > 6$, one can expect that the loop configurations with the largest loop cluster of order $L^{d/3}$ correspond to the special configuration sector in the FK Ising model.

 \par 


 \par 

\par 

	\section{Discussion}
	\label{sec:discussion}
	
In this work, we perform a large-scale Monte Carlo simulation of the  Ising model in the loop representation on high-dimensional tori for $d=5,6,7$. Our data suggest that the finite-size scaling (FSS) behaviors of the loop Ising model are simultaneously governed by the Gaussian fixed point (GFP) asymptotics and the complete-graph (CG) asymptotics. Moreover, although the loop Ising model exhibits two length scales, two configuration sectors and two scaling windows, as the Fortuin-Kasteleyn (FK) Ising model, we find that there is only one upper critical dimension $\dc=4$ for the loop Ising model, rather than two upper critical dimensions $d_c = 4, d_p = 6$ as observed in the FK Ising model. The rich FSS behavior in the loop Ising model, together with the Loop-Cluster transformation, provides an explanation to the existence of two upper critical dimensions in the FK Ising model.

It is worth noting that, for the Ising model in the  three representations, the spin representation, the FK representation, and the loop representation, there is a common upper critical dimension $d_c=4$. Above $d_c$, scaling behaviors are simultaneously governed by the CG and GFP asymptotics, which provides a unified picture for the high-dimensional Ising model.   
In the spin representation, the GFP asymptotics account for the FSS of distance-dependent observables including the short-distance behavior of the two-point correlation function and the nonzero Fourier modes of the susceptibility, etc. On the other hand, the CG asymptotics acts as the ``background”, contributing to the leading FSS behavior of the conventional macroscopic observables, such as the magnetization, energy, susceptibility, and the specific heat, etc. 
In the loop representation, the GFP and the CG asymptotics respectively describe the FSS behavior of loop clusters with radii less than and exceeding the system size $L$.
For the FK representation, the largest cluster follows the CG asymptotics for all $d>4$, but other clusters follow the GFP-Ising asymptotics for $4 < d < 6$ but follow GFP-percolation behavior for $d \geq 6$.

 \section*{Acknowledgements}
	This work has been supported by the National Natural Science Foundation of China (under Grant No. 12275263), the Innovation Program for Quantum Science and Technology (under grant No. 2021ZD0301900), the Natural Science Foundation of Fujian Province of China(under Grant No. 2023J02032).

\providecommand{\noopsort}[1]{}\providecommand{\singleletter}[1]{#1}%

\end{document}